\documentclass[12pt,preprint]{aastex}

\begin{document}
\title{Supernova Fallback:  A Possible Site for the r-Process}

\author
{Christopher L. Fryer \altaffilmark{1,2},
Herwig, Falk \altaffilmark{2},
Hungerford, Aimee\altaffilmark{4},
and F.X. Timmes \altaffilmark{1,3}
}

\altaffiltext{1}{Department of Physics, The University of Arizona,
Tucson, AZ 85721} 
\altaffiltext{2}{Theoretical Division, LANL, Los Alamos, NM 87545}
\altaffiltext{3}{Applied Physics Division, LANL, Los Alamos, NM 87545}
\altaffiltext{4}{Computer and Computation Sciences Division, LANL, Los Alamos, NM 87545}

\begin{abstract}

The conditions for the leading r-process site candidate,
neutrino-driven winds, can not be reproduced self-consistently in
current supernova models. For that reason, we investigate an alternate
model involving the mass ejected by fallback in a supernova explosion,
through hydrodynamic and nucleosynthesis calculations. The
nucleosynthetic products of this ejected material produces r-process
elements, including those in the vicinity of the elusive 3rd peak at
mass number 195. Trans-iron element production beyond the second peak
is made possible by a rapid ($<1ms$) freezeout of $\alpha$ particles
which leaves behind a large nucleon (including protons!) to r-process
seed ratio. This rapid phase is followed by a relatively long
($\gtrsim 15\mathrm{ms}$) simmering phase at $\sim$ 2$\times$10$^9$ K,
which is the thermodynamic consequence of the hydrodynamic trajectory
of the turbulent flows in the fallback outburst. During the slow phase
high mass elements beyond the second peak are first made through rapid
capture of both protons and neutrons. The flow stays close to valley
of stability during this phase. After freeze-out of protons the
remaining neutrons cause a shift out to short-lived isotopes as is
typical for the r-process.  A low electron fraction isn't required in
this model, however, the detailed final distribution is sensitive to
the electron fraction. Our simulations suggest that supernova fallback
is a viable alternative scenario for the r-process.
\end{abstract}


\section{Introduction}

Production mechanisms for elements heavier than iron, by fast
(r-process) and slow (s-process) captures of neutrons, have been known
for a long time (Cameron 1957, Burbidge et al. 1957), yet
finding conditions in Nature and understanding the physics that allow
these mechanisms to robustly operate has been more elusive.  The
search for an r-process production site has proven especially
difficult.  The supernova wind model (see Qian \& Woosley 1996 for a
detailed description), invoking the neutrino-driven wind produced by
the cooling proto-neutron star formed in a core-collapse supernovae,
is the best-studied r-process mechanism.  However, the wind scenario
generally seem to require uncommon conditions (e.g. $> 2$\,M$_\odot$
neutron star; Argast et al. 2004; Suzuki \& Nagataki 2005) to achieve
a reasonable r-process signature.

Difficulties with the wind mechanism have led to investigations about
other possible sites for the r-process; the best-studied being the
coalescence of two neutron stars (Freiburghaus 1999).  Unfortunately,
the low event rate of merging neutron stars appears to rule out binary
coalescence as a primary production site (Argast et al. 2004).  Most
other mechanisms have been based on the suggestions by Qian \& Woosley
(1996) from changes in the neutrino/nuclear physics (e.g.  neutrino
oscillations - Qian et al. 1993) to magnetic fields (e.g. Suzuki \&
Nagataki 2005).  In this paper we explore supernova fallback, yet
another of Qian \& Woosley (1996).

After the launch of a supernova explosion, some of the material
initially ejected in the blast can decelerate and fall back onto the
proto-neutron star.  This ``fallback'' material was initially proposed
by Colgate (1971) who argued that a rarefaction wave would catch the
shock and the lack of pressure support would cause the shock to fall
back onto the neutron star.  Essentially, the material with ejecta
velocities below the escape velocity will fall back onto the neutron
star.  Shigeyama, Nomoto, \& Hashimoto (1988) and Woosley (1989)
proposed a scenario where fallback occurs when the shock decelerates
as it moves through the star.  This deceleration sends a reverse shock
through the star, decelerating the innermost material and causing it
to accrete.  This mechanism is strongest when the shock hits the
hydrogen envelop, attaining its most dramatic deceleration.
Simulations show that fallback occurs early (Fryer et al. 1999;
MacFadyen et al. 2001), establishing that sub-escape velocity ejecta
dominates fallback material (Fryer \& Kalogera 2001).  Indeed, Fryer
\& Heger (2000) found fallback in a disk in the first second after the
launch of the explosion.

Although simulations of supernova explosions suggest that fallback
occurs in all simulations, this is not yet the prevailing view among
core-collapse theorists.  Fallback is strongest when the explosion is
weak and it may be that only for weak supernovae, M $\ge$
20\,M$_\odot$, that the fallback mechanism can work (Fryer 1999).
This does not preclude fallback as an r-process source!  In fact,
Argast et al. (2004) found that narrow ranges of massive stars
can explain the entire r-process abundance pattern.

In this paper, we present the first calculations of the r-process
based on hydrodynamic simulations of fallback, the ejection of fallback material (\S 2), and detailed
nucleosynthesis calculations of the ejecta (\S 3).  Without tuning our
initial conditions, we find that fallback leads to ejecta that carry
abundance signatures characteristic for r-process, thereby
demonstrating that ejecta from the fallback of material onto neutron
stars is a viable r-process site.

\section{Ejecta from Fallback}

We have modeled the fallback material and ejection of a fraction of
this material using the 2-dimensional Smooth Particle Hydrodynamics
code described in Fryer et al. (1996).  The neutron star is modeled as
a 1.4\,M$_\odot$, 10\,km hard surface emitting neutrinos.  This code
includes an equation of state valid from densities below
1\,g\,cm$^{-3}$ up to nuclear densities and a flux-limited diffusion
neutrino transport scheme for 3 neutrino species (Herant et al. 1994).
The mass and entropy of the ejecta depend upon a range of assumptions
for the initial conditions: neutrino luminosity and energy and the
accretion rate, angular momentum and composition of the infalling flow.

For this simulation, we modelled an early time ($<50$\,s) fallback and
added a neutrino flux of 2$\times$10$^{51}$ erg s$^{-1}$ in electron
neutrinos (with a mean energy of 10\,MeV), and 1.6$\times$10$^{51}$ 
erg s$^{-1}$ in electron anti-neutrinos (with a mean energy of
15\,MeV).  The results presented here do not depend sensitively on
this choice of neutrino emission, as neutrino absorption is not the
dominant force driving mass ejection. We also do not incorporate
neutrino absorption in the presented nucleosynthesis calculations.
That is, we do not use the electron fractions determined in the
hydrodynamics calculation in the post-process nucleosynthesis
calculation, but instead set the electron fraction to a constant
value: Y$_e$= 0.5 for our standard calculation.

The accretion rates from current fallback simulations predict a range
of mass inflow rates.  Piston-driven explosions using the
1-dimensional stellar evolution code KEPLER (Fryer et al. 1999) found
accretion rates ranged from nearly 10$^{4}$-10$^{5}$ M$_{\odot}$
y$^{-1}$ over a brief time (accreting roughly 0.1\,M$_\odot$) for low
mass progenitors down to 10$^{4}$ M$_{\odot}$ y$^{-1}$ for an extended
period for a 25\,M$_\odot$ star (accreting over 1\,M$_{\odot}$).
MacFadyen et al. (2001) studied a 25\,M$_{\odot}$ progenitor with a
range of explosion energies with accretion rates between 10$^{3}$ and
10$^{4}$ M$_{\odot}$ y$^{-1}$ for strong explosions producing neutron
star remnants to fallback rates as high as 10$^{6}$ M$_{\odot}$
y$^{-1}$ that ultimately produce black holes.  For our simulations, we
used the fairly normal 10$^{4}$ M$_{\odot}$ y$^{-1}$ value,
representing a snapshot in time of the fallback in a supernova
explosion.

The angular momentum in stellar cores, and hence fallback, is still
quite uncertain.  Heger et al. (2000, 2004) find the angular momentum
in the core at the time of collapse ranges from 10$^{15 - 17}$ cm$^2$
s$^{-1}$.  In our calculation, we have assumed an angular momentum at
the low end of this range 10$^{15}$ cm$^2$ s$^{-1}$.  With such a low
angular momentum, the accreting material does not form a centrifugally
supported disk.  However, the angular momentum does affect the flow,
as we shall see at the end of this section.  For high angular momenta,
the infalling material will form a disk and disk outflows will drive
most of the ejecta.

Lastly, we had to choose the composition of the fallback material.  It
has long been believed that the ejecta from core-collapse will be
neutron rich (Arnett \& Truran 1970) and many of the succesful
explosion models have ejected neutron-rich (Y$_e<$0.5) material
(e.g. Herant et al. 1994).  Indeed, it was in an effort to remove these
neutron rich ejecta that Colgate (1971) began to study fallback.  Some
recent calculations (e.g. Pruet et al. 2005) have found that the
neutrinos reset the electron fraction leading to ejecta that are proton
rich (Y$_e >$0.5).  To reset the electron fraction, the neutrino-driven
wind must play a dominant role in driving the explosion (more likely
in weak explosions with considerable fallback).  We have used two
initial compositions, one with Y$_e$=0.5 (all $^{56}$Ni), more
consistent with the most recent results, and one with $Y_e=0.49$
($^{56}$Ni with some $^{52}$Fe), closer to past results.

\begin{figure}
\plotone{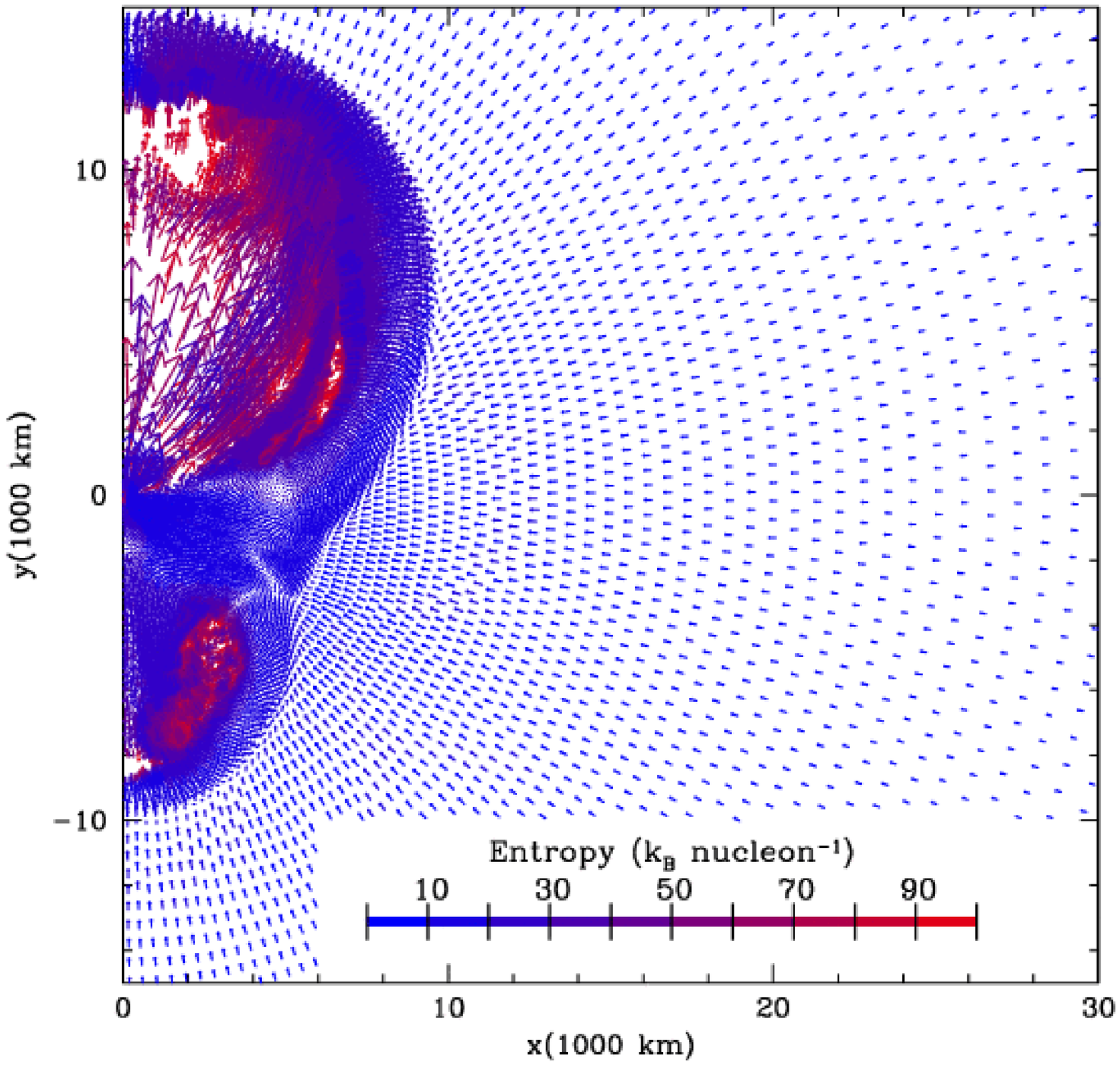}
\caption{Accretion and mass ejecta in supernova fallback 0.82s after
initial fallback.  Each point denotes particle position in the x-y
plane of the 2-dimensional smooth particle hydrodynamics simulation
where the y-axis is the axis of symmetry (set to the angular momentum
axis of the fallback).  The vectors denote velocity direction and
magnitude (length of vector).  The colors show entropy in units of
Boltzmann constant per nucleon.}
\label{fig:hydro}
\end{figure}

With this 2-dimensional smooth particle hydrodynamics code and these
initial conditions, we followed the evolution of the fallback for
3.5\,s.  The results for our Y$_e$=0.5 simulation at 0.82\,s are shown
in figure \ref{fig:hydro}.  Material crashes down onto the neutron
star and is shocked, in some cases to entropies above 250 k$_B$
nucleon$^{-1}$.  Some of the material bubbles up and is driven off the
neutron star by the energy released from the accreting material.
Recall that material accreting onto the neutron star releases roughly
10$^{20}$ erg g$^{-1}$. A small amount of accreting material can drive
off 100 times its mass with ejecta velocities of 10,000 km s$^{-1}$ if
the cooling is inefficient (because of the high temperature dependence
of neutrino emission, this is often the case) and if you have some
means of transporting energy out (e.g. viscous heating).  This can
occur because the infalling matter is only marginally bound.  The
potential energy is converted to kinetic and ultimately thermal energy
during the infall, but the infalling matter remains only marginally
bound throughout the infall (if cooling is inefficient, energy is
conserved).  If it can get a small amount of energy from its
neighboring matter, it can become unbound.  In our simulation, roughly
25\% of all our accreting material is ejected with velocities greater
than the escape velocity.  The efficiency at which material is ejected
depends on the angular momentum of the infalling material, neutrino
cooling and neutrino heating.  That there is ejecta is not a surprise,
and the nature of this ejecta has been studied over a range of
conditions.  In the limit of high angular momentum and an absorbing
boundary (i.e. black hole accretion disk), these outflows are
well-studied: see Blandford \& Begelman (1999) for a review.  With a
hard surface boundary, such as we would expect from our central
neutron star, we expect outflows even at low angular momenta (Fryer et
al. 1996).

The bubbling up ejecta from fallback expands and cools quickly through
adiabatic expansion.  Some matter shocked to temperatures above
10$^{10}$K can cool down to 2-4$\times$10$^9$ K on millisecond
timescales.  But as these bubbles push against the additional fallback
material, the expansion, and hence cooling, slows.  This produces a
simmering phase that is important for nucleon captures and the
r-process.  Although the rapid temperature drop and simmering phase
should be generic features of matter ejected in fallback, the exact
temperatures at which these two phases occur depends upon the initial
conditions.  Unfortunately, the yields depend sensitively on these
temperatures.  The dependencies on these different physical effects
will be studied in detail in a later paper.  For this paper, we focus
on testing whether fallback ejecta can provide a viable r-process.  It
is the material with velocities greater than the escape velocity that
we study in detail with our nuclear network.

\section{Nucleosynthesis of Ejecta from Fallback}

The thermodynamic histories of 6617 particles ($\sim$ 25\% of the
particles in the collapse) that reached escape velocity were
post-processed with a 3304 isotope network.  The temperature and
density histories of this material can be highly non-monotonic and is
not sufficiently described by a simple adiabatic expansion or wind
ansatz.  The reaction rates were taken from experiment whenever
possible, from detailed shell-model based calculations (Fisker et
al. 2001) and from Hauser-Feshbach calculations (Rauscher \&
Thielemann 2000).  Separation energies were taken from a combination
of experiment (Audi \& Wapstra 1995), the Hartree-Fock Coulomb
displacement calculations of Brown et al. (2002) and theoretical
estimates (M\"oller et al. 1995).  The influence of thermal effects on
weak decays was estimated from the Fuller et al. (1982). The reaction
network was integrated with the semi-implicit, variable order
algorithms described in Timmes (1999).  Our reaction network features
a nuclear statistical equilibrium (NSE) calculation to determine the
abundances when the temperature exceeds 10$^{10}$ K.  This increases
computational efficiency by over an order-of-magnitude, while
producing very little discontinuity upon either entering or leaving
NSE.

\begin{figure}
\plotone{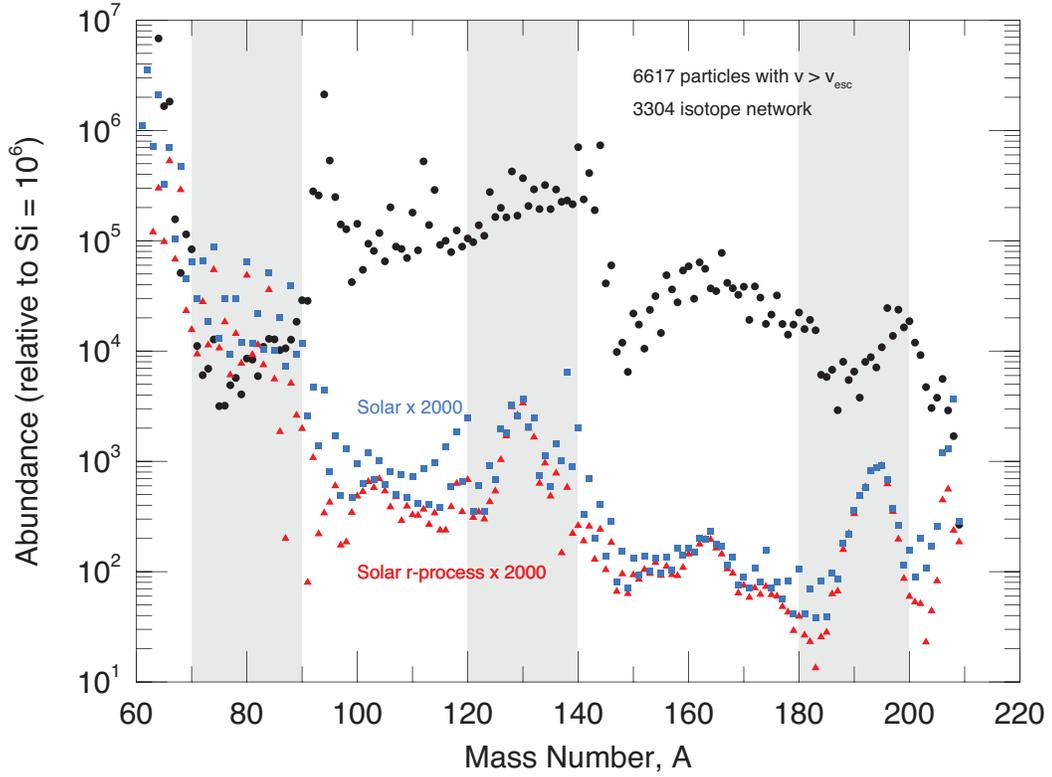}
\caption{Abundance pattern of the stable isotopes for all the
particles that achieved escape velocity (black circles).  The calculation
assumes all particles had an initial composition characterized by
Y$_e$=0.5.  The x-axis is the atomic mass number.  The y-axis is the
logarithm of the model abundance, nomalized to an elemental silicon
abundance of 10$^6$. The Anders \& Grevesse (1989) total solar abundance
pattern is shown as the blue squares with r-process yields shown 
as red triangles.}
\label{fig:rclassic_ye0p5}
\end{figure}

Figure \ref{fig:rclassic_ye0p5} shows the stable isotope distribution
attained by the 6617 particles that reach escape velocity as the black
circles. This calculation assumed that every particle had an initial
composition characterized by Y$_e$=0.5.  For nearly all of the
trajectories of interest the temperature goes above 10$^{10}$ K and
memory of the exact initial composition is forgotten except for its
Y$_e$. Particles that populate the region around the A=195 peak are
mainly produced by a three-step operation. A rapid freezeout as a result
of rapid expansion causes a persistent disequilibrium between free
nucleons and abundant alpha particles (Meyer 2002) followed by a
relatively long simmering phase $\sim$ 2$\times$10$^9$ K with proton
and neutron mass fractions $\sim$ 3$\times$10$^{-3}$.

\begin{figure}

\plottwo{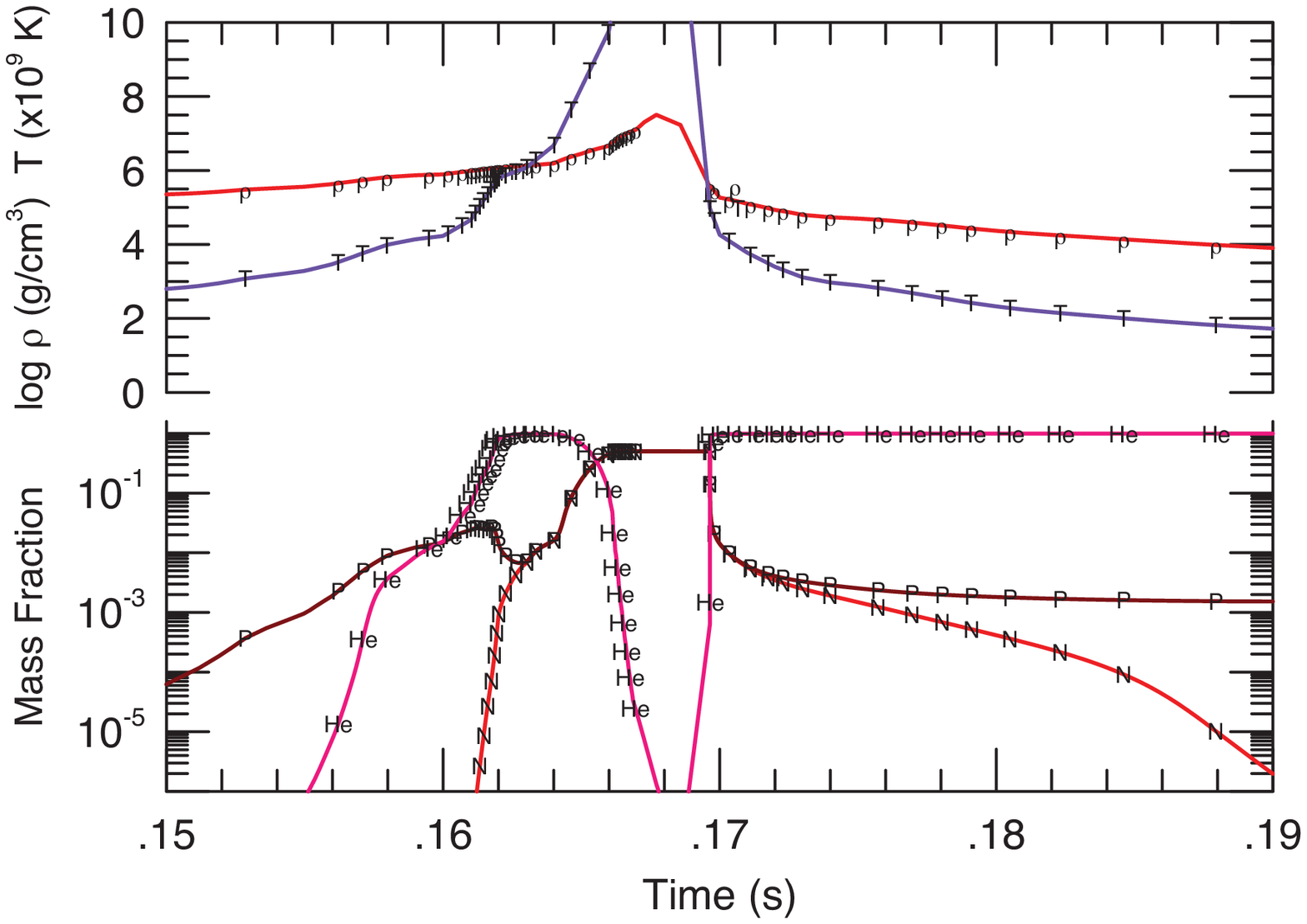}{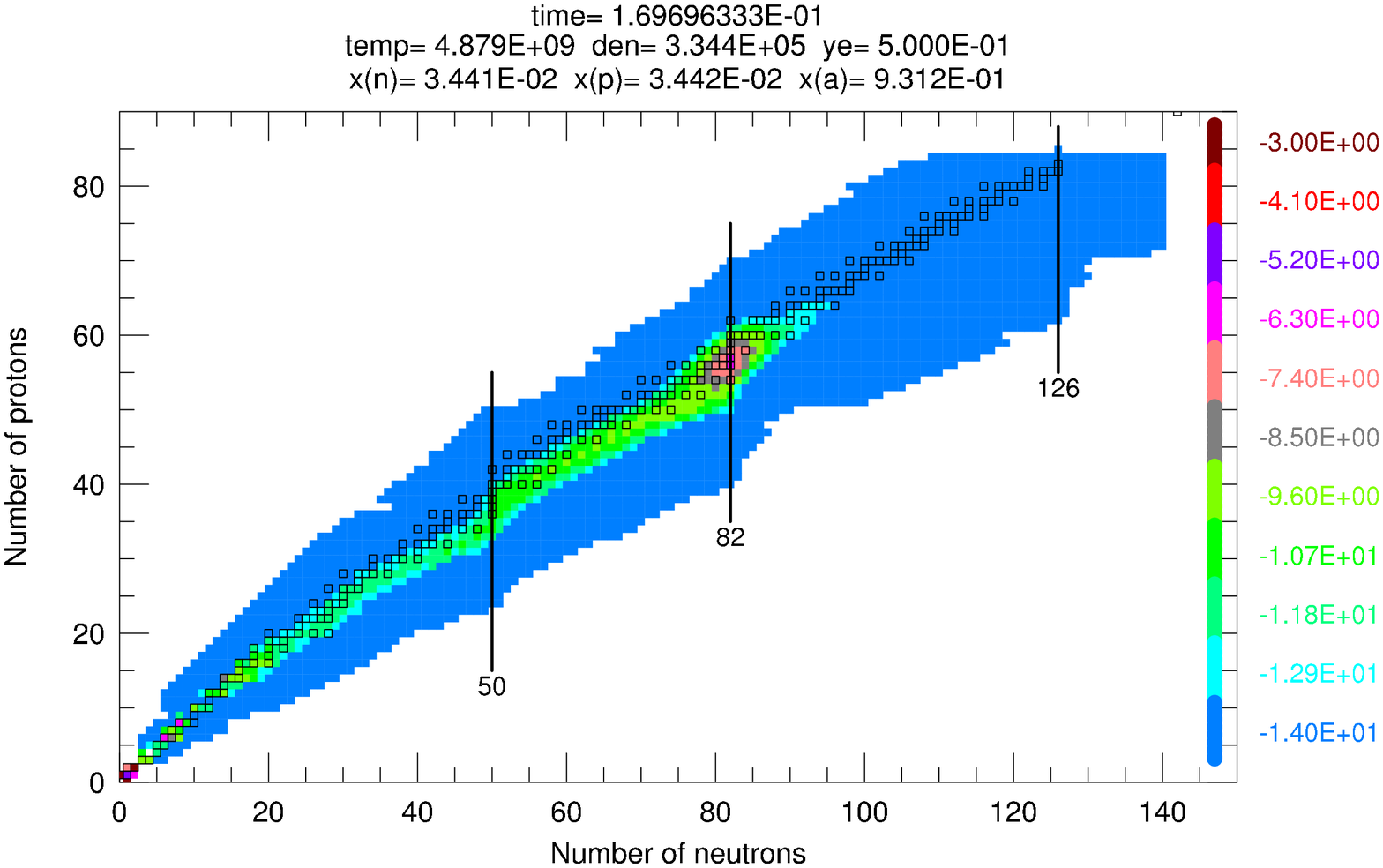}

\vspace{0.3in}

\plottwo{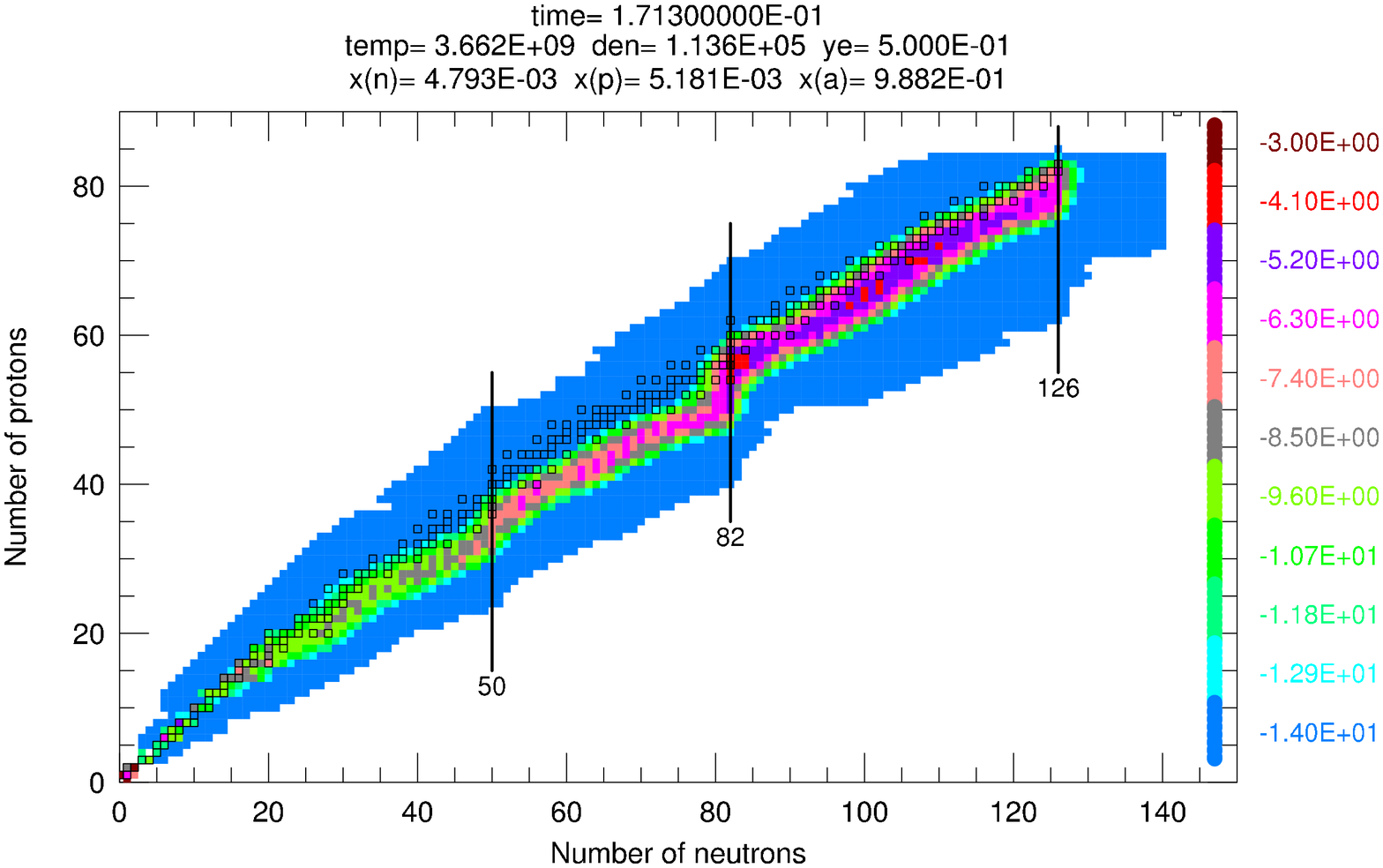}{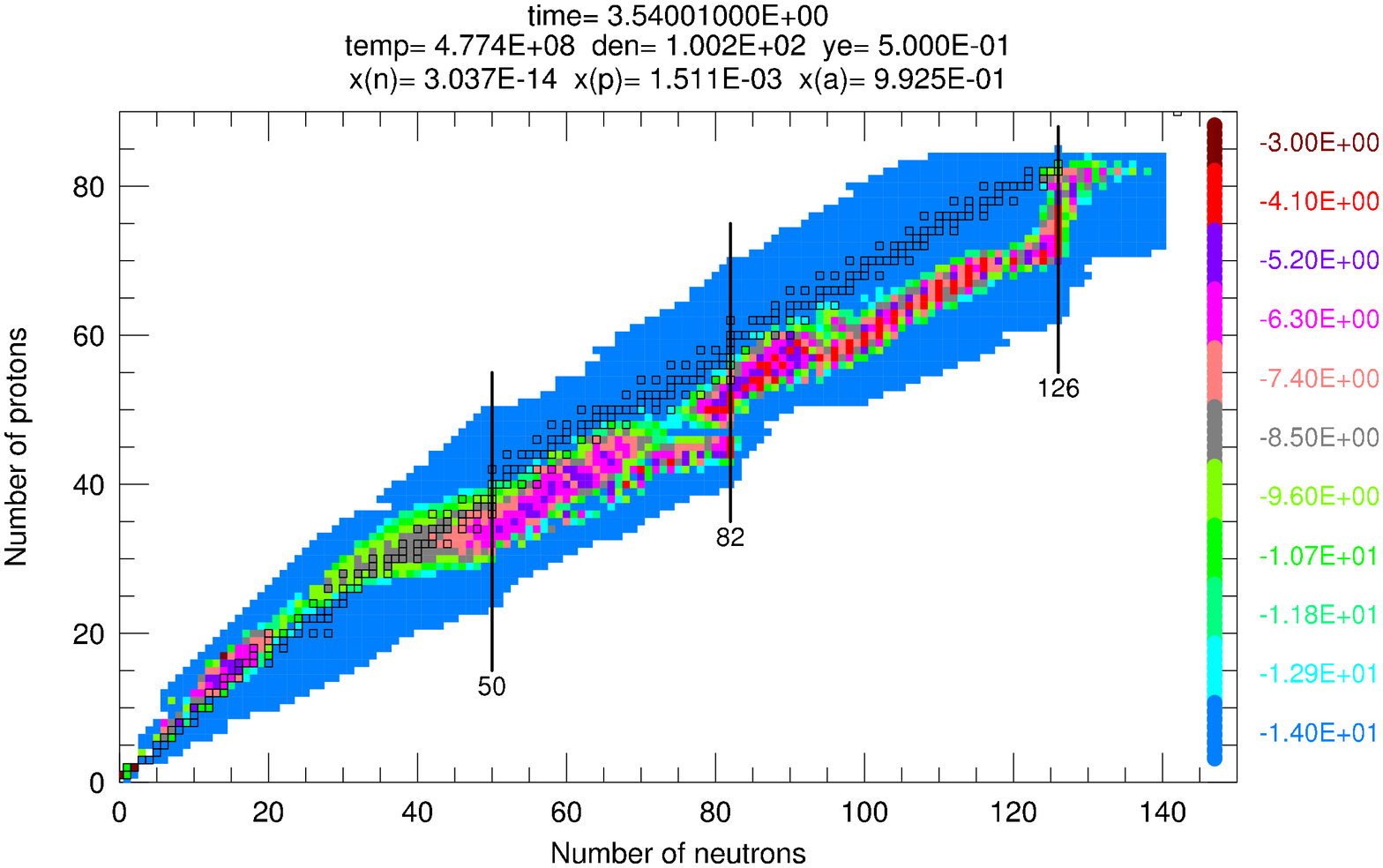}

\caption{Evolution of the temperature, density, and neutron,
proton and alpha abundances for one specific ejected particle.  The sharp
drop in temperature makes the neutrons and protons fall into
disequilibrium with the alpha particles, leaving free neutrons and
protons to capture onto heavy elements.  We also show the element
abundance at 3 different snapshots in time: 0.170\,s 0.171\,s and
3.5\,s.  At 0.17\,s, the material has just fallen from high
temperatures down to a few times $10^9$K.  The heavy elements center
around the neutron numbers of 82.  In the next 10\,ms, rapid neutron
and proton capture drive this material up to neutron numbers of 126.  
After this time, the temperature is too cool to allow proton capture, 
but neutron capture continues to drive the elements neutron rich, 
producing an r-process-like signature.}
\label{fig:time}
\end{figure}

The flows that populate the region near A=195 stay initially close to
the valley of beta stability.  The nuclei mass is driven up past the
neutron-closed shell at N=82 by an irregular, alternating succession
of neutron and proton captures. This phase after $\alpha$ freeze out
and before proton freeze out can be thought of as rapid proton and
neutron capture process, hence the ``rpn-process''.  Figure
\ref{fig:time} shows the trajectory of a single particle that produces
elements near the A=195 peak along with snapshots of its isotopic
abundances. The matter of this particle is shocked to temperatures
well beyond 1\,MeV.  Panel three (lower left) shows the
nucleosynthesis situation towards the end of this rpn-process phase. Then, at $t=0.172\mathrm{s}$ for this
particular particle, protons freeze out, and the remaining neutrons
are captured on a time scale of a few ms, driving the flow out to
short half lives that are characteristic for the r-process.

Although the flows stay initially relatively close to the valley of
beta stability, this material synthesizes r-process and not s-process
elements. Figure \ref{fig:rclassic_ye0p5} shows the solar abundance
distribution, and the solar r-process component.  Our distribution
does not show some typical signatures of canonical s-process,
including large abundances of $^{138}$Ba and $^{208}$Pb.  Our models
produce several elements, including Ba, Pb and Hg, which have a
significant s-process contribution in the solar distribution. However,
in our distribution the s-only isotopes are absent, which is another
indication that our mechanism will not lead to s-process
distribution. In addition, our global Ba/Eu ratio in the total ejected
material is $\mathrm{[Ba/Eu]}=-0.2$, which in observed stars would be
taken as a clear indication of a r-process signature.

Although the existence of a A$\sim$195 peak in our simulations shows
the potential of fallback ejecta as an r-process site, our current
simulation is far from reproducing the solar r-process signature.  The
yields from our simulation produce peaks that are wider and at
slightly higher masses than observed.  One reason for this could well
be the uncertainties in the beta decay rates.  Engel et al. (1999)
have found that more accurate calculations of the beta decay rates
lead to shorter half-lives, which cause the third peak to occur at
lower mass. Such a shift might make our peaks more nearly match the
observed data (see also Farouqi et al. 2005).  Jordan \& Meyer (2004)
have more generally altered nuclear rates and found that the exact
yield depends sensitively on this rates.  So the differences between
our results and the observed data may be resolved, at least partially,
by uncertainties in the nuclear rates.  The differences can also be
resolved (as Meyer 2002 has already pointed out), by altering the
exact value of the electron fraction.  1\% variations in the electron
fraction can change the yield from a clear r-process signature
(Y$_e$=0.495) to a proton-rich yield (Y$_e$=0.505).

Our simulations of the ejecta from supernova fallback suggest the
potential of this site to produce heavy elements in supernovae.  Rapid
cooling followed by a simmering phase allows rapid proton and neutron
captures with a final neutron burst to produce heavy elements, even
with $Y_e \approx 0.5$.  But we are far from reproducing the exact
r-process yields.  To determine if fallback is Nature's chosen site for
r-process element production, we must include the time evolution of
the electron fraction for each individual particle.  Future work will
also study the dependence of the yields on the fallback structure:
accretion rate, angular momentum (which can drastically change the
nature of the outflows), neutrino luminosity of the neutron star and
initial electron fraction.  A range of these values can occur in one
supernova explosion. By studying the nucleosynthesis we may be able to
derive more complete and realistic r-process yields for a given
supernova explosion.

{\bf Acknowledgments} We thank Al Cameron, Andy Davis, George
Jordan, John Cowan, Brad Meyer, Hendrik Schatz, and Jim Truran for
pragmatic discussions.  This work was funded in part under the
auspices of the U.S.\ Dept.\ of Energy, and supported by its contract
W-7405-ENG-36 to Los Alamos National Laboratory, by a DOE SciDAC grant
DE-FC02-01ER41176.

{}

\newpage

\clearpage

\end{document}